\documentclass[11pt]{paper}
\usepackage{amssymb,latexsym, amsmath}
\usepackage{amscd}

\usepackage[dvips]{graphicx}
 \newtheorem{thm}{Theorem}[section]
 
 \newtheorem{lem}{Lemma}[section]

\numberwithin{equation}{section}

\renewcommand{\setminus}{\smallsetminus}

\font\rrm=wncyr10%
\font\rit=wncyi10

\newcommand{\R}{\mathbb{R}}
\newcommand{\N}{\mathbb{N}}

\DeclareMathOperator{\diag}{\mathrm{diag}}

\title{HEISENBERG MODEL IN PSEUDO--EUCLIDEAN SPACES}

\author{ Bo\v zidar Jovanovi\'c}

\begin{document}

\maketitle

\leftline{\small Mathematical Institute SANU, Serbian Academy of
Sciences and Arts} \leftline{\small Kneza Mihaila 36, 11000
Belgrade, Serbia} \leftline{\small E-mail: bozaj@mi.sanu.ac.rs}

\begin{abstract}
We construct  analogues of the classical Heisenberg spin chain
model (or the discrete Neumann system) on pseudo--spheres and
light--like cones in the pseudo--Euclidean spaces and show their
complete Hamiltonian integrability. Further, we prove that the
Heisenberg model on a light--like cone leads to a new example of
integrable discrete contact system.
\end{abstract}

\keywords{Discrete Hamiltonian and contact systems, the Lax
representation, complete integrability. {\bf MSC 2010:} 37J55,
37J35}

\section{Introduction}

A pseudo--Euclidean space $E^{k,l}$ of signature $(k,l)$,
$k,l\in\N,\, k+l=n$, is the space $\R^{n}$ endowed with the scalar
product
\[
\langle x,y\rangle = \sum_{i=1}^k x_iy_i - \sum_{i=k+1}^{n}
x_iy_i\quad (x,y\in\R^{n}).
\]

A vector $x\in E^{k,l}$ is called \emph{space--like},
\emph{time--like}, \emph{light--like}, if $\langle x,x\rangle$ is
positive, negative, or zero, respectively. Denote by
$(\cdot,\cdot)$ the Euclidean inner product in $\R^{n}$ and let
$$
E=\diag(\tau_1,\dots,\tau_n)=\mbox{diag}(1,\dots,1,-1,\dots,-1),
$$
where $k$ diagonal elements are equal to 1 and $l$ to $-1$. Then
$\langle x,y\rangle=(Ex,y)$, for all $x,y\in\R^{n}$.

We consider a discrete system on a pseudo--sphere and on the
light-like cone given by
$$
S_c^{n-1}=\{\langle q,q\rangle=(q,Eq)=c\},
$$
$c = \pm 1$ and, respectively, $c = 0$. The system is  defined by
the action functional
$$
S[\mathbf q]=\sum_k \mathbf{L}(q_k,q_{k+1}),
$$
with the discrete Lagrangian
$$
\mathbf L(q_k,q_{k+1})=\langle q_{k},J q_{k+1}\rangle=(q_k, EJ
q_{k+1}),
$$
where $\mathbf q=(q_k), \, k\in\mathbb Z$ is a sequence of points
on $S^{n-1}_c$ and $J=\mbox{diag}(J_1,\dots,J_{n})$, $J_i\ne 0,
\,i=1,\dots,n$.  The equations of the stationary configuration
have the form
\begin{equation}\label{neumann0}
\frac{\partial \mathbf L(q_k,q_{k+1})}{\partial q_k}+
\frac{\partial \mathbf L(q_{k-1},q_{k})}{\partial q_k}=\lambda_k E
q_k, \qquad k\in\mathbb Z,
\end{equation}
that is,
\begin{equation}\label{neumann}
q_{k+1}+q_{k-1}=\lambda_k J^{-1} q_k, \qquad k\in\mathbb Z,
\end{equation}
where the multipliers are determined by the constraints
\begin{equation}\label{prva}
c=\langle q_{k+1},q_{k+1}\rangle=\langle
q_{k-1},q_{k-1}\rangle-2\lambda_k\langle J^{-1}
q_k,q_{k-1}\rangle+\lambda^2_k\langle J^{-2} q_k,q_k\rangle.
\end{equation}

Thus either $\lambda_k=0$ or, in the case $ \langle J^{-2}
q_k,q_k\rangle\ne 0$, we have another solution
\begin{equation}\label{lambda}
\lambda_k=2\langle J^{-1} q_k,q_{k-1}\rangle/\langle J^{-2}
q_k,q_k\rangle.
\end{equation}
We consider the dynamics given by the second expression, which is
defined outside the singular set $\langle J^{-2} q_k,q_k
\rangle=0$.

For $n=3$ and the Euclidean case, the functional defines the
energy of a classical spin chain in the Heisenberg model
\cite{Veselov, Ves3}. Also, in the Euclidean case for arbitrary
$n$ the equations represent a discretisation of the classical
Neumann system \cite{MV, Su}. So we refer to \eqref{neumann},
\eqref{lambda} as a {\it Heisenberg model}, or a {\it discrete
Neumann system} in a pseudo--Euclidean space $E^{k,l}$.

We present a matrix Lax representation of the mapping, show that
it is symplectic, and prove its complete integrability. Recently,
a related elliptical billiard problem in pseudo--Euclidean spaces
was studied in \cite{KT, KT2, DR}. The light--like billiard flow
provides a natural example of a discrete contact completely
integrable system \cite{BM, KT2, Jov}. We shall prove that the
Heisenberg model on a light--like cone leads to an integrable
contact system as well.

\section{Description of the dynamics}

\paragraph{Symplectic description.}
Let
$$
P_c=S^{n-1}_c \times S^{n-1}_c\subset E^{k,l}\times
E^{k,l}(q,Q):\quad \phi_1=\langle q,q\rangle=c, \quad
\phi_2=\langle Q,Q\rangle=c.
$$
The equations \eqref{neumann}, \eqref{lambda} determine the
mapping
$$
\Phi: \quad P_c\longrightarrow P_c
$$
($\Phi(q_{k-1},q_k)=(q_k,q_{k+1})$), defined
outside the singular set $\langle J^{-2} Q,Q \rangle=0$. Let
$$
\alpha=\frac{\partial\mathbf{L}(q,Q)}{\partial q} dq=EJQdq=\sum_i
\tau_iJ_i Q_i dq_i\,.
$$

The flow of $\Phi$ preserves 2-form  $\Omega=d\alpha=\sum_i \tau_i
J_i dQ_i\wedge dq_i$ (see Veselov \cite{Veselov, Ves3})
$$
\Phi^*\Omega=\Omega.
$$

Namely, $\Phi$ preserves a form $\omega$ if the restriction of
$\pi_1^*\omega-\pi_2^*\omega$ to the associated graph
$$
\Gamma_\Phi=\{ (q_1,Q_1,q_2,Q_2)\in P_c\times
P_c\,\vert\,Q_1=q_2,\,\frac{\partial \mathbf L(q_2,Q_2)}{\partial
q}+ \frac{\partial \mathbf L(q_1,Q_1)}{\partial Q}=\lambda E Q_1\}
$$ equals zero, where $\pi_1$ and
$\pi_2$ are projections to the first and the second factor of
$P_c\times P_c$ and $\lambda=2\langle J^{-1}
q_1,Q_1\rangle/\langle J^{-2} Q_1,Q_1\rangle$. At $\Gamma_\Phi$,
we have
\begin{eqnarray}
\pi_1^*\alpha-\pi_2^*\alpha &=& \frac{\partial\mathbf{L}(q_1,Q_1)}{\partial q} dq_1- \frac{\partial\mathbf{L}(q_2,Q_2)}{\partial q} dq_2\nonumber\\
&=& \frac{\partial\mathbf{L}(q_1,Q_1)}{\partial q}
dq_1+\frac{\partial\mathbf{L}(q_1,Q_1)}{\partial Q} dQ_1-\lambda E Q_1dQ_1\nonumber\\
&=& d(\pi_1^*\mathbf L)-\frac12\lambda d(\langle
Q_1,Q_1\rangle)\label{grafik}
\end{eqnarray}
implying
$$
\pi_1^*\Omega-\pi_2^*\Omega=-\frac12 d\lambda \wedge d(\langle
Q,Q\rangle)=0, \quad  \text{at}\quad  \Gamma_\Phi.
$$

The form $\Omega$ is symplectic on $E^{k,l}\times E^{k,l}$. Let
$\{\cdot,\cdot\}_\Omega$ be the corresponding Poisson bracket. We
have
$$
\{\phi_1,\phi_2\}_\Omega=-\frac{\partial\phi_1}{\partial Q}\cdot
EJ^{-1}\frac{\partial\phi_2}{\partial
q}+\frac{\partial\phi_1}{\partial q}\cdot
EJ^{-1}\frac{\partial\phi_2}{\partial Q}= 4\langle
q,J^{-1}Q\rangle.
$$

Therefore, the subvariety
$$
P_c^*=P_c\setminus\{  \langle q,J^{-1}Q\rangle=0  \}
$$
is symplectic. Moreover, $P_c^*$ is $\Phi$--invariant. Indeed,
alternatively to \eqref{prva}, the Lagrange multiplier $\lambda_k$
can be found from the expression $\langle q_{k-1},q_{k-1}\rangle
=c$, provided $\langle q_{k},q_{k}\rangle =\langle
q_{k+1},q_{k+1}\rangle =c$. This leads to $ c=c-2\lambda_k\langle
J^{-1} q_k,q_{k+1}\rangle+\lambda^2_k\langle J^{-2}
q_k,q_k\rangle$ and the Lagrange multiplier equals
\begin{equation}\label{lambda2}
\lambda_k=2\langle J^{-1} q_{k+1},q_{k}\rangle/\langle J^{-2}
q_k,q_k\rangle.
\end{equation}
By combining \eqref{lambda} and \eqref{lambda2} we get that
\begin{equation}\label{K}
K(q,Q)=\langle J^{-1} Q,q \rangle=(J^{-1} E Q,q)
\end{equation}
is the first integral of the system and $P_c^*$ is
$\Phi$--invariant.

The induced Poisson bracket on $(P_c^*,\Omega)$ can be described
by the Dirac construction (e.g., \cite{Moser}):
\begin{equation}
\{f_1,f_2\}_\Omega^D=\{f_1,f_2\}_\Omega -\frac{
\{\phi_1,f_1\}_\Omega\{\phi_2,f_2\}_\Omega-\{\phi_2,f_1\}_\Omega\{\phi_1,f_2\}_\Omega}{\{\phi_1,\phi_2\}_\Omega}.
\label{Dirac_bracket}
\end{equation}

\paragraph{Light--like cone and contact description.}
In the light--like case, multiplying \eqref{neumann} by $q_k$, we
obtain the additional invariant relation
\begin{equation}\label{light}
\langle Jq_{k-1},q_k\rangle+\langle J q_k,q_{k+1}\rangle=0,
\end{equation}
which gives that $\mathbf L^2=\langle Jq,Q\rangle^2$ is also the
integral of the mapping $\Phi$. Consider the corresponding
$(2n-3)$--dimensional invariant variety
\begin{equation}\label{mk}
M_\kappa=\{(q,Q)\in P_0^*\,\vert\, \mathbf L(q,Q)=\langle
q,JQ\rangle=\pm\kappa\}=M_{\kappa,+} \bigcup M_{\kappa,-},\quad
\kappa \ge 0.
\end{equation}

\begin{thm}
The restriction of 1-form $\alpha$ to $M_\kappa$ is a contact form
for $\kappa>0$. Moreover, $\Phi$ is a contact transformation
\begin{equation}\label{kontaktna}
\Phi^*\alpha=\alpha
\end{equation}
that interchanges the components of $M_{\kappa,+}$ and
$M_{\kappa,-}$ of $M_\kappa$.
\end{thm}

The restriction of $\alpha$ to $M_\kappa$ is contact if $\alpha\ne
0$ and the restriction of $\Omega=d\alpha$  to the horizontal
distribution
$$
\mathcal H=\ker\alpha=\{\xi\in T_{(q,Q)} M_\kappa\,\vert\,
\alpha(\xi)=0\}
$$
is nondegenerate.

Let ${\tilde X}_\mathbf L$ be the Hamiltonian vector field of the
discrete Lagrangian $\mathbf L(q,Q)$ on $(P_0^*,\Omega)$. It
appears that the Hamiltonian vector field $X_\mathbf L=(q,-Q)$ of
$\mathbf L$ on $(E^{k,l}\times E^{k,l},\Omega)$ is tangent to
$P_0^*$, and therefore,
$$
{\tilde X}_{\mathbf L}=X_\mathbf L\vert_{P_0^*}=(q,-Q).
$$

From \eqref{mk} and the definition of the Hamiltonian vector field
${\tilde X}_\mathbf L$ ($i_{{\tilde X}_\mathbf L}\Omega=-d\mathbf
L\vert_{P_0^*}$), the kernel of $\Omega$ restricted to $M_\kappa$
is proportional to ${\tilde X}_\mathbf L$. Further,
\begin{equation}\label{ham}
\alpha({\tilde X}_{\mathbf L})=(EJq,Q)=\mathbf L=\pm\kappa,\qquad
(q,Q)\in M_{\kappa,\pm}
\end{equation}
implies that $\ker d\alpha \cap \mathcal H=0$ and the restriction
of $d\alpha$ to $\mathcal H$ is nondegenerate, for $\kappa>0$.

Next, by restricting \eqref{grafik} to $M_\kappa\times M_\kappa$,
and using the fact that both, the restrictions of $d\mathbf L$ and
$d(\langle Q_1,Q_1\rangle)$ to $M_\kappa$ are equal to zero, we
obtain \eqref{kontaktna}.

\section{Integrability}
Define
$$
F=\diag(1,\dots,1,i,\dots, i),
$$
where the first $k$ components are equal to 1, and the last $n-k$
components are equal to the imaginary unit $i$ ($F^2=E$).

Motivated by the Moser--Veselov Lax matrix representation for the
Heisenberg system in the Euclidean case \cite{MV}, we obtain the
following statement.

\begin{thm}
The equations \eqref{neumann} implies the matrix equation
$$
L_{k+1}(\lambda)=A_k(\lambda)L_k(\lambda)A_k^{-1}(\lambda),
$$
where
$$
L_{k}(\lambda)=J^2+\lambda Fq_{k-1}\wedge FJ q_k-\lambda^2 c\cdot
Fq_{k-1}\otimes Fq_{k-1}, \quad A_k(\lambda)=J-\lambda Fq_k\otimes
Fq_{k-1}.
$$
Like in \cite{MV}, we have the matrix factorization
$L_k=A_k^T(-\lambda)A_k(\lambda)$. Therefore, the discrete Lax
representation can be seen also as interchanging of
$A_k$--matrixes:
$$
L_{k+1}=A_{k+1}^T(-\lambda)A_{k+1}(\lambda)=A_k(\lambda)A_k^T(-\lambda).
$$
\end{thm}

Note that in the case of a Heisenberg model on a light--like cone
($c=0$), the $L$--matrix is linear in $\lambda$.

If $J^2_j\ne J^2_i$, the integrals of the mapping $\Phi$ obtained
from the matrix representation can be written in the form
\begin{equation}\label{int-neumann}
f_i(q,Q)=c\cdot \tau_i q_i^2+\sum_{j\neq
i}\frac{\tau_i\tau_j(J_jQ_jq_i-q_jJ_iQ_i)^2}{ J^2_i- J^2_j}, \quad
i=1,\dots,n.
\end{equation}

\begin{lem}
\label{relacije} The integrals \eqref{K}, \eqref{int-neumann} on
$P_c$ are related by
\begin{eqnarray}
&& \sum_i f_i \equiv c^2, \label{rel}\\
&& \sum_i \frac{1}{J_i^2} f_i \equiv K^2,\label{rel2}
\end{eqnarray}
and, for $c=0$, we have the additional relation
\begin{equation}
\sum_i J_i^2 f_i \equiv -\mathbf L^2.\label{rel3}
\end{equation}
\end{lem}

By direct calculations, one can prove that the integrals
\eqref{int-neumann} commute on $(P_c^*,\Omega)$:
\begin{equation}\label{commuting}
\{f_i,f_j\}_\Omega^D=0, \qquad i,j=1,\dots,n
\end{equation}
and that the only relation among them is \eqref{rel}.

\begin{thm}
The Heisenberg model is a completely integrable discrete
Hamiltonian system on $(P_c^*,\Omega)$, $c=\pm 1, 0$.
\end{thm}

We proceed with the light--like case and consider the mapping
$\Phi$ restricted to the contact manifold $(M_\kappa,\alpha)$.
Recall that a vector field $Y$ is contact if it preserves the
horizontal distribution $\mathcal H$, i.e., $\mathcal
L_Y\alpha=\lambda\alpha$, for some smooth function $\lambda$. The
distinguish contact vector field is the Reeb vector field $Z$,
uniquely defined by
\begin{equation*}
i_Z\alpha=1, \qquad i_Z d\alpha=0.
\end{equation*}

From \eqref{ham}, the Reeb vector field reads
$$
Z\vert_{(q,Q)}=\pm\frac{1}{\kappa}(q,-Q), \qquad (q,Q)\in
M_{\kappa,\pm}.
$$

Since the Lie derivatives of the integrals \eqref{int-neumann}
along $X_\mathbf L$, for $c=0$, are equal to zero
$$
\mathcal L_{X_\mathbf L} f_i=\{f_i,\mathbf L\}_\Omega=0 \qquad
i=1,\dots,n,
$$
we obtain
\begin{equation}\label{reeb}
\mathcal L_{Z} f_i=0 \qquad \Longleftrightarrow \qquad
[Z,Y_{f_i}]=0
\end{equation}
where $Y_{f_i}$ are the contact Hamiltonian vector fields
 with Hamiltonians $f_i$,
$i=1,\dots,n$. \footnote{The mapping $\Psi: Y \mapsto f=i_Y\alpha$
establish the isomorphism between the vector spaces of contact
vector fields and smooth functions on $M_\kappa$. The vector field
$Y_f=\Psi^{-1}(f)$ is the {\it contact Hamiltonian vector field}
with the Hamiltonian function $f$. In particular, the Hamiltonian
of the Reeb vector field is $f\equiv 1$. Further, $\Psi$ is a Lie
algebra isomorphism: $Y_{[f,g]}=[Y_f,Y_g],$ where on the right
hand side we have the usual Lie bracket of vector fields, while on
the left hand side we have the {\it Jacobi bracket} on
$C^\infty(M_\kappa)$ defined by $ [f,g]=\mathcal L_{Y_f}g- g
\mathcal L_Z f$ (e.g., see \cite{LM}).}

From \eqref{commuting} and \eqref{reeb}, using the theorem on {\it
isoenergetic integrability} (for more details, see \cite{JJ2}), we
get
\begin{equation}\label{commuting2}
[f_i,f_j]=0 \qquad \Longleftrightarrow \qquad [Y_{f_i},Y_{f_j}]=0
\qquad  i,j=1,\dots,n.
\end{equation}

Apart from \eqref{rel}, on $M_\kappa$ the integrals
\eqref{int-neumann} have the additional relation $ \sum_i J_i^2
f_i\equiv-\kappa^2$ (see \eqref{rel3}), and there are $n-2$
independent functions among them.

Thus, the mapping $\Phi$
 is a completely integrable contact transformation (see \cite{KT2, Jov}). The
conditions \eqref{reeb}, \eqref{commuting2},
 ensure that $M_\kappa$ is almost everywhere foliated on
$(n-1)$--dimensional invariant manifolds, regular level sets of
integrals $f_1,\dots,f_n$. The associated distribution $\mathcal
F$ is pre--Legendrian (or co--Legendrian)\footnote{$\mathcal F$ is
transversal to $\mathcal H$ and $\mathcal G=\mathcal F\cap
\mathcal H$ is a maximal integrable (Legendrian) distribution of
$\mathcal H$.} and it is generated by the commuting contact vector
fields $Z,Y_{f_1},\dots,Y_{f_n}$ \cite{KT2}.

\begin{thm}
The Heisenberg model on the light--like cone in a
pseudo--Euclidean space $E^{k,l}$ is completely integrable contact
system on $(M_\kappa,\alpha)$, $\kappa>0$.
\end{thm}

\subsection*{Acknowledgments}
I would like to thank the referee for useful suggestions. The
research was supported by the Serbian Ministry of Science Project
174020, Geometry and Topology of Manifolds, Classical Mechanics
and Integrable Dynamical Systems.


\begin{thebibliography}{84}

\bibitem{BM} A. Banyaga and P, Molino,  \emph{G\' eom\' etrie des
formes de contact compl\'etement int\'egrables de type torique},
S\'eminare Gaston Darboux, Montpellier (1991-92), 1-25. See also:
Complete Integrability in Contact Geometry, Penn State preprint PM
197, 1996.


\bibitem{DR}
V. Dragovi\'c and M. Radnovi\'c, \emph{Ellipsoidal billiards in
pseudo-euclidean spaces and relativistic quadrics}, Adv. Math.
\textbf{231} (2012), 1173-1201, arXiv:1108.4552 [math.AG].

\bibitem{Jov} B. Jovanovi\' c, \emph{Noncommutative integrability and action angle variables in contact
geometry}, Journal of Symplectic Geometry,  \textbf{10} (2012),
535--562, arXiv:1103.3611 [math.SG].

\bibitem{JJ2} B. Jovanovi\' c, V. Jovanovi\' c, \emph{Contact flows
and integrable systems}, arXiv:1212.2918.

\bibitem{LM} P. Libermann, C. Marle, \emph{Symplectic Geometry,
Analytical Mechanics}, Riedel, Dordrecht, 1987.


\bibitem{KT}
B. Khesin and S. Tabachnikov, \emph{Pseudo-Riemannian geodesics
and billiards}, Adv. Math.  \textbf{221} (2009), 1364--1396,
arXiv:math/0608620 [math.DG] 

\bibitem{KT2} B. Khesin, S. Tabachnikov, \emph{Contact complete
integrability, Regular and Chaotic Dynamics},  \textbf{15} (2010)
 504–-520, arXiv:0910.0375 [math.SG].

\bibitem{Moser} {J. Moser},
\emph{Geometry of quadric and spectral theory}. In: Chern
Symposium 1979, Berlin--Heidelberg--New York, 147--188, 1980.

\bibitem{MV}  J. Moser, A. P.  Veselov,
\emph{Discrete versions of some classical integrable systems and
factorization of matrix polynomials}, {Comm. Math. Phys.} {\bf
139} (1991) 217--243.

\bibitem{Su} Yu. B. Suris
 \emph{The problem of integrable discretization: Hamiltonian
 approach},
Progress in Mathematics, {\bf 219}. Birkhäuser Verlag, Basel,
2003.

\bibitem{Veselov} {\rrm A. P. Veselov, {\rit Integriruemye sistemy s diskretnym vremenem
i raznostnye operatory},  Funkc. analiz i ego prilozh.}
\textbf{22}(2) (1988), 1--13 (Russian); English translation:

A. P. Veselov, \emph{Integrable discrete--time systems and and
difference operators}, Funct. Anal. Appl. {\bf 22} (1988), 83--94

\bibitem{Ves3}  {\rrm A. P. Veselov, {\rit Integriruemye otobrazheniya},
Uspehi Mat. Nauk,} \textbf{46}(5) (1991), 3--45 (Russian); English
translation:

A. P. Veselov, \emph{Integrable maps}, Russ. Math. Surv.
\textbf{46} (5) (1991) 1--51.
\end{thebibliography}
\end{document}